\documentclass[fleqn,usenatbib]{mnras}

\usepackage{newtxtext,newtxmath}
\usepackage[T1]{fontenc}
\usepackage{graphicx}	
\usepackage{amsmath}	
\usepackage{subcaption}
\usepackage{scalerel}

\def\Msun{\, M_{\odot}}

\DeclareRobustCommand{\VAN}[3]{#2}
\let\VANthebibliography\thebibliography
\def\thebibliography{\DeclareRobustCommand{\VAN}[3]{##3}\VANthebibliography}

\title[Spatial and orbital planes of the Milky Way satellites]{Spatial and orbital planes of the Milky Way satellites: unusual but consistent with $\Lambda$CDM}

\author[Pham, Kravtsov \& Manwadkar]{
Khanh Pham$^1$, Andrey Kravtsov$^{1,2,3}\thanks{E-mail: kravtsov@uchicago.edu}$ and Viraj Manwadkar$^1$
\\
$^{1}$Department of Astronomy  \& Astrophysics, The University of Chicago, Chicago, IL 60637 USA\\
$^{2}$Kavli Institute for Cosmological Physics, The University of Chicago, Chicago, IL 60637 USA\\
$^{3}$Enrico Fermi Institute, The University of Chicago, Chicago, IL 60637 USA
}

\date{Accepted XXX. Received YYY; in original form ZZZ}

\pubyear{2022}

\begin{document}
\label{firstpage}
\pagerange{\pageref{firstpage}--\pageref{lastpage}}
\maketitle

\begin{abstract}
We examine the spatial distribution and orbital pole correlations of satellites in a suite of zoom-in high-resolution dissipationless simulations of Miky Way (MW) sized haloes. We use the measured distribution to estimate the incidence of satellite configurations as flattened and as correlated in their orbital pole distribution as satellite system of the Milky Way. We confirm that this incidence is sensitive to the radial distribution of subhaloes and thereby to the processes that affect it, such as artificial disruption due to numerical effects and disruption due to the central disk. Controlling for the resolution effects and bracketing the effects of the disk, we find that the MW satellite system is somewhat unusual (at the $\approx 2-3\sigma$ level) but is statistically consistent with the $\Lambda$CDM model, in general agreement with results and conclusions of other recent studies.
\end{abstract}

\begin{keywords}
galaxies: evolution, galaxies: formation, galaxies: dwarf, galaxies: haloes, galaxy: star formation
\end{keywords}



\section{Introduction}
\label{sec:intro}

Studies of dwarf galaxies are at the current frontier of studies of galaxy formation and properties of dwarf galaxy satellites are an important testing ground for galaxy formation models and properties of dark matter (DM) particles \citep[see, e.g.,][for a review]{bullock_boylankolchin17}. One of the puzzling properties of observed dwarf satellite galaxies of the Milky Way (MW) and a number of nearby massive galaxies is anisotropy of their spatial and orbital configurations. Spatial distributions of satellites tend to be flattened and in some  systems significant fraction of satellites exhibits signs of coherent rotation with strong correlation of orbital poles \citep[see, e.g.,][for recent reviews]{Pawlowski.2018,Pawlowski.2021}.

The observed satellite configurations are often considered to be a challenge for the $\Lambda$CDM model \citep[e.g.,][]{kroupa_etal05,Pawlowski.2021,Mueller.etal.2021,Perivolaropoulos.Skara.2022}. Although multiple comparisons with predictions of the $\Lambda$CDM model have been carried out over the past two decades
(see Section 2.4 of \citealt{bullock_boylankolchin17}, Section 1 of \citealt{Pawlowski.etal.2019}, or Section 2.2.3 of \citealt{Pawlowski.2021} for reviews),
review of recent theoretical studies  shows that there is no consensus on whether observed satellite configurations are statistically consistent with $\Lambda$CDM model because the estimates of the theoretical prediction for their incidence rate vary widely. For example,  \citet{pawlowski_kroupa20} argued that the incidence rate of the satellite systems with spatial and orbital anisotropy similar to the Milky Way is negligibly small, while \citet[][]{Samuel.etal.2021}, \citet[][]{sawala_etal22}, and \citet{Foerster.etal.2022} conclude that the incidence of such configurations in cosmological simulations is sufficiently high to be statistically consistent with observations. 

In this study we use high-resolution zoom-in simulations of the MW-sized haloes to examine spatial distribution of satellites and correlation of their orbital poles. Specifically, we construct distributions of the thickness of the satellite planes, their median galacto-centric radii, and dispersion of their orbital poles.  We use these distributions to estimate the incidence of $\Lambda$CDM satellite configurations as flattened and as correlated in their orbital poles as bright satellites of the Milky Way. 

Although flattened satellite systems have been found around several nearby galaxies, the available information about these systems is rather different in each galaxy and statistics used to characterized them are often very different as well. In this study we focus on the MW because its census of bright satellites is 1) the most complete and 2) the best characterized in terms of 3D spatial distribution and orbital properties.  

Given that recent studies cited above reached opposite conclusions about incidence of the MW-like satellite systems, we believe that it is warranted to focus on the best studied case of the MW specifically. Additionally, certain lessons related to numerical subhalo disruption and baryonic effects that arise in this and other recent studies have relevance for comparisons with anisotropic satellite configurations around other nearby galaxies. 

This paper is organized as follows. We describe the simulations used and the modelling of the luminous Milky Way satellite systems in Section~\ref{sec:modelling}. We present comparisons of model predictions with observed flattening and orbital pole anisotropy of the MW satellites in Section~\ref{sec:results}, discuss our results and compare to other relevant studies in Section \ref{sec:discussion}, and summarize our results and conclusions in Section~\ref{sec:summary}. We present comparison of radial distributions of subhaloes in the ELVIS and Caterpillar simulation suites and inferences about numerical convergence in the Appendix~\ref{app:elvis_cat_rad_comp}.

\section{Modelling Milky Way satellite system}
\label{sec:modelling}

To model the observed population of dwarf satellite galaxies around the Milky Way, we used tracks of haloes and subhaloes in the ELVIS \citep{GarrisonKimmel.etal.2014} and Caterpillar \citep{Griffen.etal.2016} suites of $N$-body simulations of the MW-sized haloes described in the two subsections below. The two suites are used to gauge robustness of our results and effects of resolution. 

The mass assembly histories of subhaloes around MW-sized haloes are used as a baseline for the \texttt{GRUMPY} galaxy evolution framework to model evolution of baryonic component of galaxies they host.
This framework is described and tested against a wide range of observations of local dwarf galaxies in \citet[][]{Kravtsov.Manwadkar.2022} and was shown to be capable in reproducing luminosity function and radial distribution of the Milky Way satellites \citep{Manwadkar.Kravtsov.2022}. Here we use the model to predict luminosities of satellite galaxies and select the 11 brightest ones to compared with 11 brightest MW satellites. 
We summarize the main features of the model in Subsection~\ref{sec:grumpy} below. Finally, in subsection~\ref{sec:subhalo_disrupt} we describe the model accounting for subhalo disruption due to tidal forces exerted by the central Milky Way disk that is lacking in the dissipationless $N$-body simulations.

\subsection{ELVIS simulation suite}
\label{sec:elvis}

The ELVIS suite of high-resolution simulations \citep{GarrisonKimmel.etal.2014} 
consists of ``zoom-in'' high-resolution regions of $\sim 1 $ Mpc around 24 isolated MW-sized haloes and twelve pairs of MW-sized haloes separated by $\sim 800$ kpc using GADGET-2 and GADGET-3 codes \citep{Springel.2005} in a box of size 70.4 Mpc. Simulations were initialized using  MUSIC code \citep{hahn_abel11} and run in the flat $\Lambda$CDM model with $\Omega_{\rm m} = 0.266$, $\Omega_\Lambda =
0.734$, $n_s = 0.963$, $\sigma_8 = 0.801$, and $h = H_0/100=0.71$. The corresponding particle mass within the high-resolution zoom-in region is $m_{\rm p}=1.9\times  10^5\, M_\odot$, while the
Plummer-equivalent force softening was held fixed in physical coordinates at $z<9$ at 141 pc. 

We use evolution tracks of haloes and subhaloes that exist at $z=0$ in and around Milky Way-sized halo from each zoom-in region. The tracks were extracted from simulations using the Consistent Trees code \citep{Behroozi.etal.2013} and consist of  several halo properties, such as its virial mass, scale radius, maximum circular velocity, etc., measured at a series of redshifts from the first epoch at which progenitors are identified to $z=0$. 

\subsection{Caterpillar simulation suite}
\label{sec:caterpillar}
 
We also use halo catalogs and halo mass accretion histories (MAHs) from the Caterpillar suite \citep[][]{Griffen.etal.2016} of zoom-in simulations of MW-sized haloes.\footnote{\url{https://www.caterpillarproject.org}}  The Caterpillar suite consists of simulations of 32 MW-like host haloes\footnote{A total of 35 MW-like haloes were simulated, but a few of them are not used here as they are contaminated by low resolution particles \citep[see][]{Griffen.etal.2016}.}, which have been re-simulated at three different resolution levels, namely LX12, LX13 and LX14, with LX14 being the highest resolution of the three \citep[see][for details]{Griffen.etal.2016}. This allows checks for convergence of the results. In particular, in the Appendix~\ref{app:elvis_cat_rad_comp} we demonstrate that the LX14 resolution is higher than the resolution of ELVIS simulations. 
 
The haloes were identified using the modified version of the Rockstar halo finder and the Consistent Trees Code \citep{Behroozi.etal.2013}, with modification intoduced to improve recovery of mass subhaloes with high initial fraction of unbound particles \citet[see discussion in Section 2.5 of][]{Griffen.etal.2016}. 

As shown in \citet[][see their Fig. 1]{Manwadkar.Kravtsov.2022} the subhalo peak mass function in the LX14 simulations flattens at $\mu=M_{\rm peak}/M_{\rm host} \approx 4 \times 10^{-6}$ ($M_{\rm peak} \approx 4 \times 10^{6} \Msun$) even in the innermost regions of the host ($r < 50$ kpc). This mass can be viewed as a rough threshold for resolved subhaloes and shows that massive subhaloes used in this study $M_{\rm peak}>5\times 10^8\,M_\odot$ are two orders of magnitude larger in mass than this threshold. Further resolution tests are presented in the Appendix~\ref{app:elvis_cat_rad_comp}. 

\begin{figure*}
    \centering
    {\includegraphics[width=0.49\textwidth]{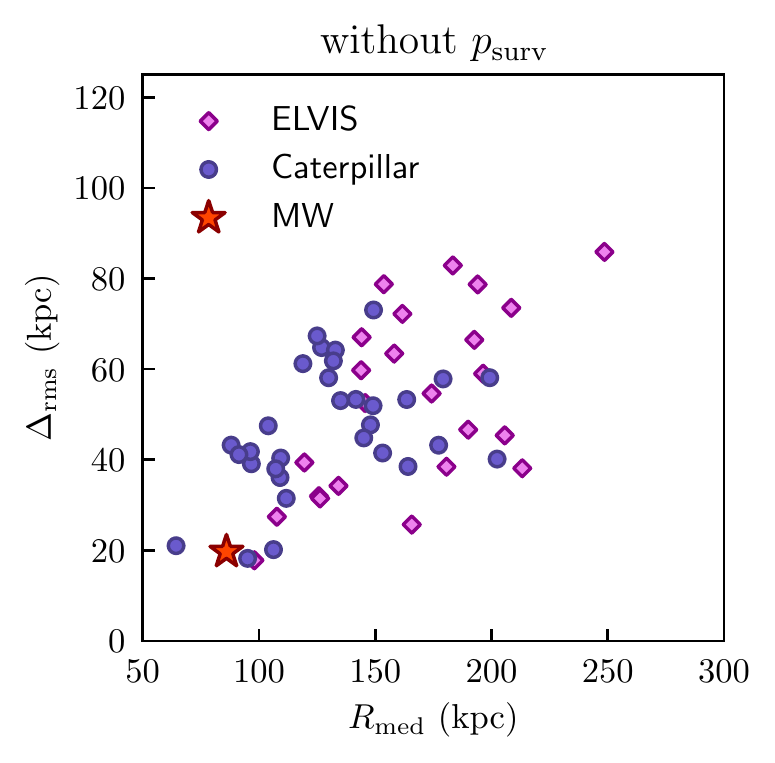}
    \includegraphics[width=0.49\textwidth]{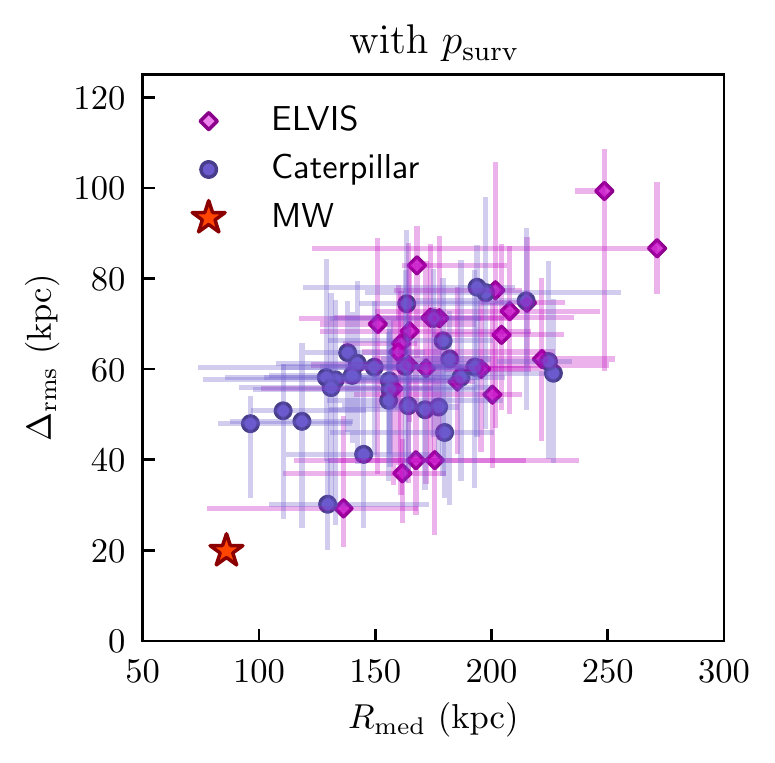}}
    \caption{Dispersion around the best fit plane ($\Delta_{\rm rms}$) versus the median distance of satellites to the host halo center ($R_{\rm med}$). {\it The left panel\/} shows these quantities for the 11 brightest satellites for each host in the isolated ELVIS and Caterpillar simulation suites, where additional tidal disruption due to central disk is not accounted for. {\it The right panel\/} shows a similar distribution but after accounting for additional disruption due to central disk using the model described in Section~\ref{sec:subhalo_disrupt}. The uncertainties show 95\%-range estimating using many bootstrapped realizations of the survival probability of subhalos in each host and estimating $R_{\rm med}$ and $\Delta_{\rm rms}$ using 11 brightest surviving satellites for each realization. }
    \label{fig:drms_rmed}
\end{figure*}

\subsection{The GRUMPY galaxy evolution model}
\label{sec:grumpy}

The \texttt{GRUMPY} (Galaxy formation with RegUlator Model in PYthon) is a regulator-type galaxy formation framework \citep[e.g.,][]{Krumholz.Dekel.2012,Lilly.etal.2013,Feldmann.2013} designed to model dwarf galaxies \citep[][]{Kravtsov.Manwadkar.2022}.

The backbone of the model is the rate of change of the total gravitating mass of the halo (baryons and DM): $\dot{M}$. We use the rates $\dot{M}$ extracted from the ELVIS and Caterpillar suites of cosmological simulations described above. Specifically, we use the halo mass enclosing the density contrast of 200 times the critical density at the redshift of analysis. 
To provide a monotonic representation of the evolution of halo mass for galaxy formation model and to smooth spurious mass fluctuations due to subhaloes, we first  approximate $\log M_{\rm 200c}$ as a function of $\log t$  using cubic spline and enforce the monotonicity condition $d\ln M_{\rm 200c}/d\ln t\geq 0$ at the times $t_i$ and approximate the resulting mass evolution again using the cubic spline.

The monotonic $M_{200\rm c}(t)$ of each subhalo is used to follow evolution of baryonic component in the galaxy it hosts. Specifically, the ISM gas mass, stellar mass, and mass in metals in these two components are followed using a system of ordinary differential equations expressing mass conservation and mass transfer between different components.
The model accounts for UV heating after reionization and associated gas accretion suppression onto small mass haloes, galactic outflows, model for gaseous disk and its size, molecular hydrogen mass, star formation, etc.  Evolution of the half-mass radius of the stellar distribution is also modelled. The galaxy model parameters used in this study are identical to those used in \citet{Manwadkar.Kravtsov.2022}. 

The model reproduces a wide range of observed properties of dwarf galaxies, including relations between gas and stellar metallicity and galaxy stellar mass, half-light radius-luminosity relation, luminosity function of the Milky Way satellites, correlation of gas and stellar masses, diversity of star formation histories, etc. \citep[see][]{Kravtsov.Manwadkar.2022,Manwadkar.Kravtsov.2022}. 

\subsection{Subhalo disruption model}
\label{sec:subhalo_disrupt}

Recent numerical studies showed that presence of the central disk decreases the overall abundance of subhaloes within the host virial radius by a factor of $\sim 0.5-0.65$ \citep[][]{Wetzel.etal.2016,Zhu.etal.2016,Garrison_Kimmel.etal.2017,Nadler.etal.2018,Kelley.et.al.2019}, although \citealt{Webb.Bovy.2020} and \citealt{Green.etal.2022} recently argued that effects are much smaller. 

To account for this additional disruption, we use the disruption model of \citet{Nadler.etal.2018} calibrated using two hosts from the FIRE-2 suite of galaxy formation simulations. The model estimates a survival probability, $p_{\rm surv}=1-p_{\rm dis}$, where $p_{\rm dis}$ is disruption probability of a subhalo given its physical distance from the host centre, scale factor at the first pericentric passage after accretion, and its virial mass, maximum circular velocity and scale factor at the time of accretion. For each halo in the simulation suites these parameters are extracted from the halo evolution tracks described above. 

With survival probabilities assigned for each subhalo in a simulation, we create 50,000 Monte Carlo realizations of surviving subhalo populations using individual $p_{\rm surv}$ values. We then select 11 brightest satellites from each realization using luminosities assigned using the GRUMPY model described in Section~\ref{sec:grumpy} above. The Monte Carlo realizations are used to compute the median values of parameters characterizing anisotropy of satellite distribution
and their percentiles to estimate uncertainty.

\section{Results}
\label{sec:results}

Although a number of different statistics can be used to characterize the anisotropy of the satellite distribution of the Milky Way, for the ease of comparison in what follows we will use the most commonly used measures of the thickness of the satellite ``plane'' and dispersion of their orbital pole directions \citep[e.g.,][]{pawlowski_kroupa20}. These statistics are defined in the following subsections. 

\subsection{Thickness of the Milky Way satellite plane}
\label{sec:drms}

The {\sl rms\/} deviation of the satellite distances from the best fit plane was used in most previous studies to characterize the thickness of the planar distribution of satellites \citep[][]{kroupa_etal05,zentner_etal05}. For $k$ satellites this quantity is defined as: 
\begin{equation}
\Delta_{\rm rms}(k) = \sqrt{\frac{\sum_{i=1}^k(\hat{ \mathbf{n}}\cdot\mathbf{x}_i)^2}{k}},
\end{equation}
where $\hat{\mathbf{n}}$ is a unit vector normal to a given plane and $\mathbf{x}_i$ is the vector from the
center of the halo to the $i$-th satellite position. 
The best fit plane is obtained by minimizing $\Delta_{\rm rms}(k)$ and this minimum value is used 
as the thickness of the satellite distribution.

\begin{figure*}
    \centering
    {\includegraphics[width=0.49\textwidth]{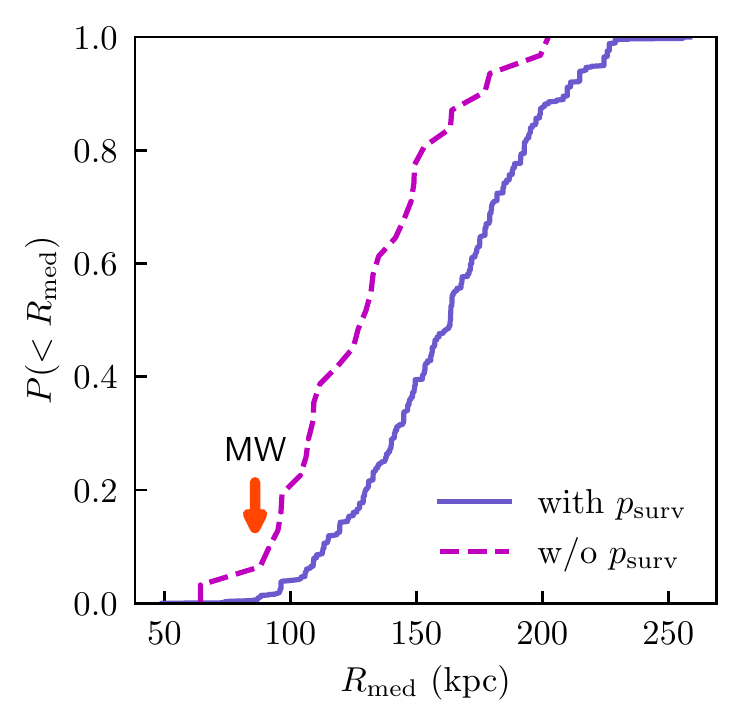}
    \includegraphics[width=0.49\textwidth]{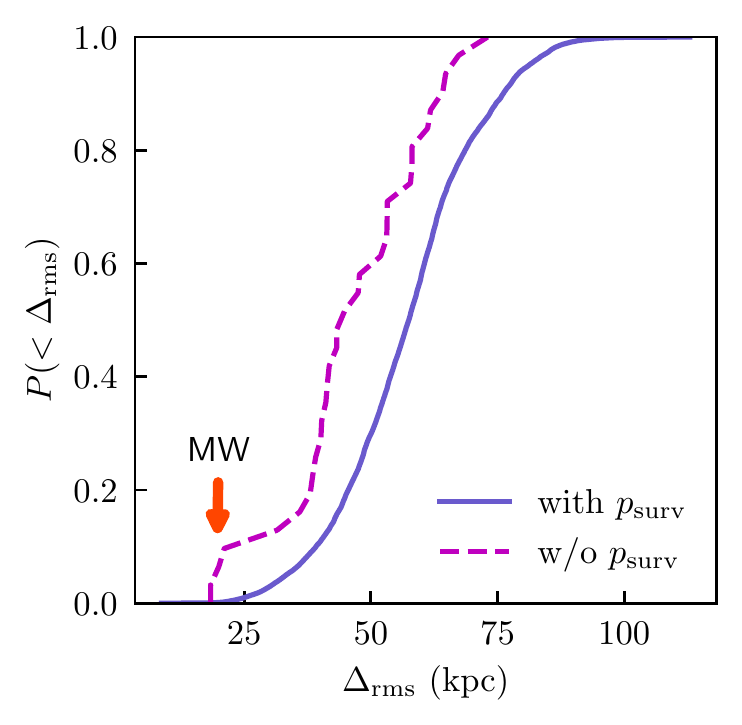}}
    \caption{(\textit{Left panel}) Cumulative distribution of the median distance of the 11 brightest satellites to the host halo center in the Caterpillar simulation suite. (\textit{Right panel}) Cumulative distribution of the dispersion around the best fit plane $\Delta_{\rm rms}$ for the 11 brightest satellites in the hosts of the Caterpillar suites. In both panels, solid blue and dashed magenta lines show the distributions with and without accounting for additional disruption due to central disk, respectively. Note that the distributions accounting for additional disruption due to the central disk are constructed using many realizations of the survival probability for each satellite. For each realization, $R_{\rm med}$ and $\Delta_{\rm rms}$ are computed for the 11 brightest surviving satellites. The orange arrow shows the corresponding quantity for the 11 brightest satellites of the MW.}
    \label{fig:rmed_drms_cdf}
\end{figure*}

\subsection{Correlation of the satellite plane thickness with median radial distance of satellites}
\label{sec:rmed_drms}

It is well known that there exists a correlation between the plane thickness $\Delta_{\rm rms}$ defined above and the concentration of the radial distribution of satellites \citep{kang_etal05,zentner_etal05}. Therefore, one can only meaningfully compare model to observations if either radial distribution of observed satellites is reasonably reproduced or if comparison is done in a scaled quantity $\Delta_{\rm rms}/R_{\rm med}$ \citep[][]{zentner_etal05} where $R_{\rm med}$ is the median distance of the satellites to the host halo center. 

This correlation is illustrated in Figure~\ref{fig:drms_rmed} which shows $\Delta_{\rm rms}$ as a function of the median distance of satellites to the host halo center for 11 brightest satellites in all of the MW-sized host haloes in the isolated ELVIS and Caterpillar simulation suites. The luminosities of satellites were computed using \texttt{GRUMPY} model described in Section~\ref{sec:grumpy}. The left panel of Figure~\ref{fig:drms_rmed} shows results for raw results of the satellite distribution, while the right panel shows results when additional tidal disruption due to the central disk was accounted for using model described in Section~\ref{sec:subhalo_disrupt} and uncertainties estimated using Monte Carlo realizations of satellite populations using disruption probability for each halo. 

The $\Delta_{\rm rms}-R_{\rm med}$ correlation is readily apparent in Figure~\ref{fig:drms_rmed} with thinnest satellite planes occurring in host haloes with most concentrated radial distribution of satellites (i.e., smallest $R_{\rm med}$ values). In fact, the Milky Way satellite system lies along this predicted correlation. When we do not account for additional disruption due to disk, three haloes in the Caterpillar suite and two haloes in the ELVIS suite have $\Delta_{\rm rms}\approx 20-30$ kpc and $R_{\rm med}$ values comparable to those of the 11 brighest satellites of the Milky Way. There is a substantial scatter in the $\Delta_{\rm rms}-R_{\rm med}$ plane, which may explain why differences in the radial distributions of satellites may not be reflected in the corresponding differences of flattening of their configuration for individual haloes \citep[][]{Samuel.etal.2021}.

Figure~\ref{fig:drms_rmed} also shows that haloes in the ELVIS suite tend to have larger $R_{\rm med}$ and $\Delta_{\rm rms}$ values compared to the Caterpillar haloes. This difference is most likely due to lower resolution of the ELVIS simulations and larger numerical effects of premature tidal disruption \citep[e.g.,][]{vandenBosch.etal.2018,vandenBosch.Ogiya.2018,Webb.Bovy.2020,Grand.etal.2021,Green.etal.2021}. As we show in Section 4.4.4 of \citet{Manwadkar.Kravtsov.2022}, these numerical effects are small for the Caterpillar suite at the LX14 resolution level and for subhaloes with peak masses $M_{\rm peak}\geq 5\times 10^8\, M_\odot$ used in this analysis (see also Figure~\ref{fig:elvis_cat_pro_comp} in Appendix~\ref{app:elvis_cat_rad_comp}). However, similar analysis carried out for the ELVIS simulations shows much larger numerical disruption effects. We present comparisons of radial distributions of subhaloes in the Caterpillar and ELVIS simulations in the Appendix~\ref{app:elvis_cat_rad_comp}, which show that radial distribution in the ELVIS simulations is comparable to that in the lower resolution LX12 Caterpillar runs, which is significantly less concentrated than the radial distribution in the higher resolution LX13 and LX14 runs. The difference is naturally explained by artificial disruption at lower resolution. It is these resolution effects that lead to a more extended radial distribution of surviving satellites and systematically larger $R_{\rm med}$ and $\Delta_{\rm rms}$ values in the ELVIS suite.

At the same time, the difference between profiles in LX13 and LX14 Caterpillar runs is quite small, which indicates that results of the LX14 should be close to convergence for the subhalo mass range we use. This conclusion is consistent with tests presented in Section 5 of \citet[][]{Webb.Bovy.2020}. For this reason, in the subsequent analyses we will solely use the Caterpillar simulation suite.

The effect of additional tidal disruption due to disk is further quantified in Figure~\ref{fig:rmed_drms_cdf} which shows the cumulative distribution functions of $R_{\rm med}$ and $\Delta_{\rm rms}$ in the Caterpillar haloes with and without accounting for additional tidal disruption due to central disk. The figure shows that without accounting for disk-induced disruption $\approx 10\%$ of the MW-hosts have satellite systems more flattened that MW, in agreement with a previous estimate of \citet{Cautun.etal.2015}. This estimate is larger than that of \citet[][]{Forrero_Romero.Arias.2018}, who used Illustris-1 and ELVIS simulations, which we attribute to the effects of artificial subhalo disruption in these simulations discusssed above.

\begin{figure*}
    \centering
    \includegraphics[width=0.49\textwidth]{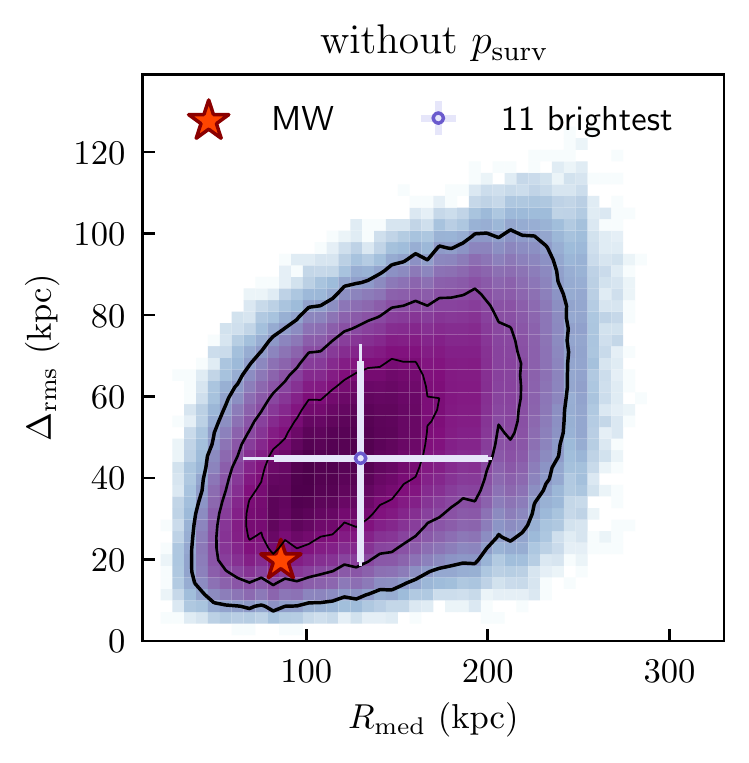}
    \includegraphics[width=0.49\textwidth]{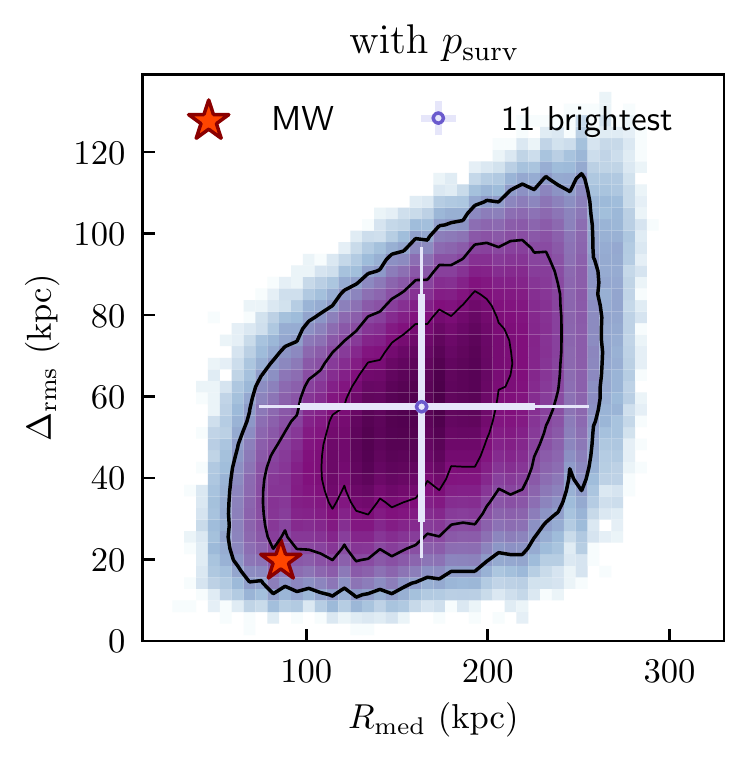}
    \caption{Dispersion around the best fit plane $\Delta_{\rm rms}$ versus the median distance of satellites to the host halo center $R_{\rm med}$ in the Caterpillar suite. The left panel shows distribution without accounting for additional disruption of satellites due to central disk, while the right panel shows distributions with such accounting. The 2D histograms show distribution of the $\Delta_{\rm rms}$ and $R_{\rm med}$ values for  randomly selected 11 subhalos with peak mass $\geq 5\times 10^8\,M_\odot$ in each Caterpillar host. The 3 contours enclose $68.27\%$, $95.45\%$, and $99.73\%$ of the samples. The lavender circle shows the median
    $\Delta_{\rm rms}$ and $R_{\rm med}$ values for the 11 brightest satellites with thick and thin error bars showing $95\%$ and $99.7\%$ percentiles of the distribution of these quantities. }
    \label{fig:drms_rmed_hist}
\end{figure*}

Figure~\ref{fig:rmed_drms_cdf} shows that both $R_{\rm med}$ and $\Delta_{\rm rms}$ are increased by the additional tidal disruption, although the magnitude of the effect is larger for $R_{\rm med}$. Incidentally, this larger effect on $R_{\rm med}$ leads to decrease of $\Delta_{\rm rms}/R_{\rm med}$ when additional tidal disruption is accounted for. 

This sizeable effect of the central baryonic disk on the radial distribution of satellites and thickness of the satellite plane significantly affects the estimate of the incidence rate of the satellite systems with thickness similar to that of the Milky Way. 
This can already be seen in Figure~\ref{fig:drms_rmed}, but the effect on the incidence rate of the MW-like satellite configurations is better quantified and in Figure~\ref{fig:drms_rmed_hist}, which shows distribution of satellite systems of the Caterpillar hosts in the $\Delta_{\rm rms}-R_{\rm med}$ plane with (right panel) and without (left panel) accounting for additional disruption of satellites due to the central disk. The 2D histograms show distribution of  randomly selected 11 subhalos with peak mass $\geq 5\times 10^8\,M_\odot$ in all Caterpillar hosts with the three contours enclosing $68.4\%$, $95.5\%$, and $99.7\%$ of the satellite systems. The lavender circle with $95.5\%$ and $99.7\%$ error bars shows $\Delta_{\rm rms}$ and $R_{\rm med}$ values for the 11 brightest satellites selected using luminosities assigned with the \texttt{GRUMPY} model.

The figure shows that without accounting for additional tidal disruption, the thickness of the MW satellite plane is within $2\sigma$ of the distribution of the Caterpillar satellite systems. 
Accounting for tidal disruption makes such systems more rare, but the observed MW satellites are still only a $2.5\sigma$ outlier. The thickness of the MW satellite plane by itself is therefore consistent with $\Lambda$CDM predictions, especially given uncertainties associated with the disk tidal disruption modelling and possible residual numerical disruption effects in the Caterpillar suite.

\begin{figure*}
    \centering
    \includegraphics[width=0.49\textwidth]{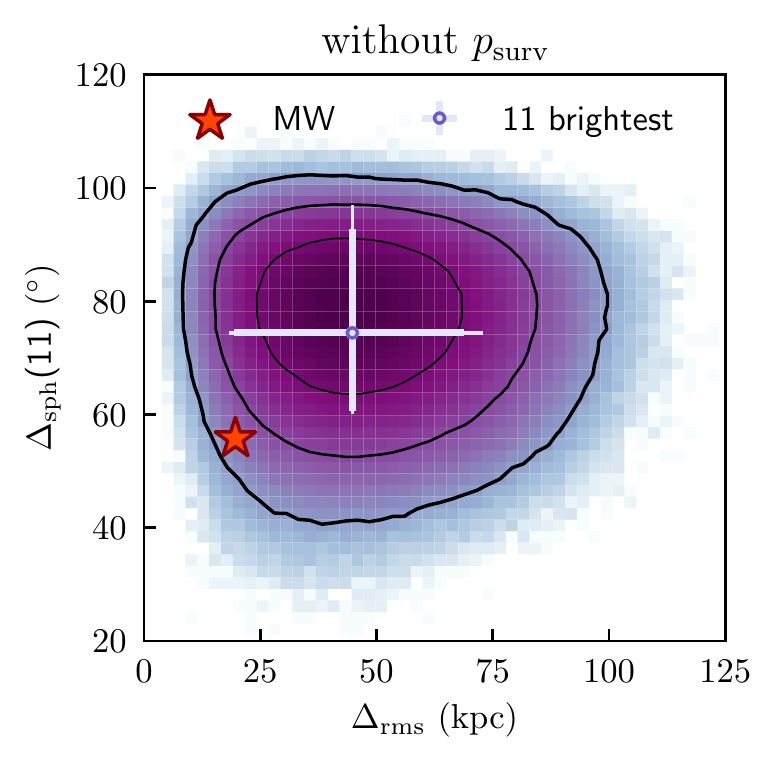}
    \includegraphics[width=0.49\textwidth]{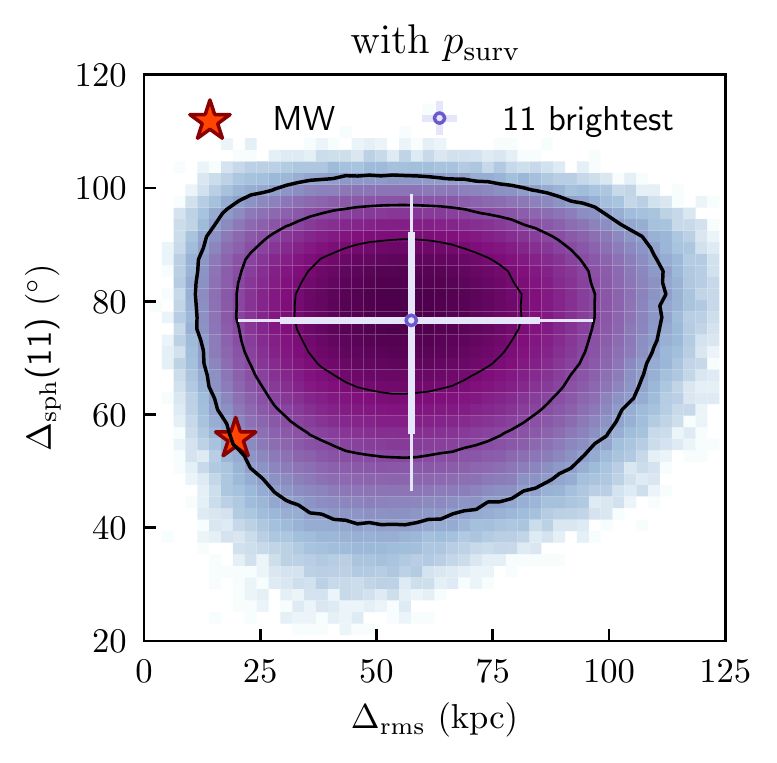}
    \caption{Spherical standard deviation of the angles of orbital poles $\Delta_{\rm sph}$ for 11 brightest satellites around the average direction of these poles vs dispersion around the best fit plane $\Delta_{\rm rms}$  for the same 11 satellites in the Caterpillar suite. {\it The left panel\/} shows distribution without accounting for additional disruption of satellites due to central disk, while {\it the right panel\/} shows distributions with such accounting. The 2D histograms show distribution of the $\Delta_{\rm sph}$ and $\Delta_{\rm rms}$  values for  randomly selected 11 subhalos with peak mass $\geq 5\times 10^8\,M_\odot$ in each Caterpillar host. The 3 contours enclose $68.27\%$, $95.45\%$, and $99.73\%$ of the samples. The lavender circle shows the median
    $\Delta_{\rm rms}$ and $R_{\rm med}$ values for the 11 brightest satellites with thick and thin error bars showing $95\%$ and $99.7\%$ percentiles of the distribution of these quantities. }
    \label{fig:dsph_drms_hist}
\end{figure*}

\begin{figure*}
    \centering
    \includegraphics[width=0.49\textwidth]{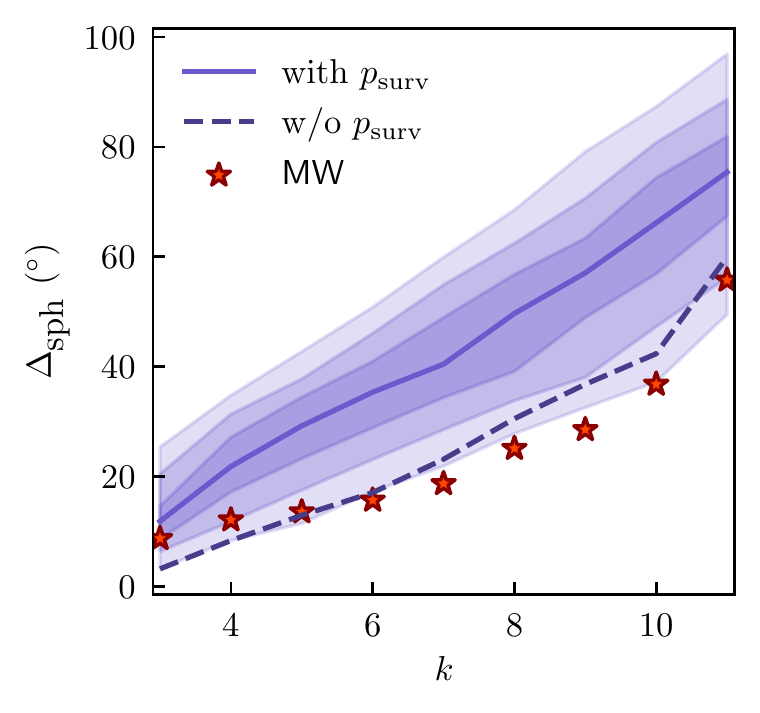}
     \includegraphics[width=0.49\textwidth]{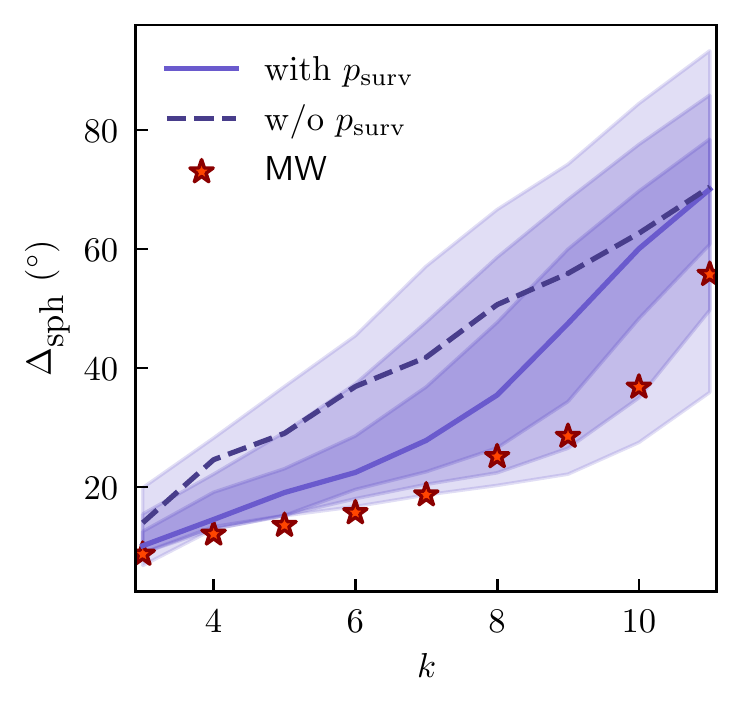}
    \caption{Spherical standard deviation of the angles of orbital poles $\Delta_{\rm sph}$ for samples of $k$ satellites ($k\in [3,11]$) for the host 17 ({\it left panel}) and host 2 ({\it right panel}) of the Caterpillar LX14 suite. The dashed lines show $\Delta_{\rm sph}(k)$ for 11 brightest satellites without accounting for additional tidal disruption due to central disk, while the solid blue line with shaded bands shows the same accounting for such disruption. In this case many Monte Carlo realizations of survival probability were done and for each realization $\Delta_{\rm sph}(k)$ was computed for 11 brightest satellites, as described in Sections~\ref{sec:subhalo_disrupt}.
    The 3 shaded bands enclose $68.27\%$, $95.45\%$, and $99.73\%$ of the values around the median value at each $k$. The red stars show the values for the MW satellites computed in a similar way. }
    \label{fig:dsphk17}
\end{figure*}

\subsection{Distribution of orbital poles}
\label{sec:poles}

To characterize the anisotropy of the satellite orbital pole directions, we use the {\sl rms\/} deviation of the satellite orbital poles positions on the unit sphere $\Delta_{\rm sph}$ \citep[][]{metz_etal07,pawlowski_etal13,pawlowski_kroupa13,pawlowski_kroupa20}. For a given number of $k$ satellites, $\Delta_{\rm sph}(k)$ is defined as
\begin{equation}
\Delta_{\rm sph}(k)=\sqrt{\frac{\sum_{i=1}^k[\arccos(\langle \mathbf{n}\rangle_p\cdot\mathbf{n}_i)]^2}{k}},
\end{equation}
where $\mathbf{n}_i$ are the vectors from the center of the unit sphere to the location of the orbital pole on the sphere of individual satellites and $\langle\mathbf{n}\rangle_p$ is the average of these vectors.

Figure~\ref{fig:dsph_drms_hist} shows the joint distribution of $\Delta_{\rm sph}(k=11)$ and $\Delta_{\rm rms}$ for satellite systems of the Caterpillar hosts with (right panel) and without (left panel) accounting for additional disruption of satellites due to central disk. The 2D histograms show distribution of  randomly selected 11 subhalos with peak mass $\geq 5\times 10^8\,M_\odot$ in all Caterpillar hosts with the three contours enclosing $68.4\%$, $95.5\%$, and $99.7\%$ of the satellite systems. The lavender circle with $95.5\%$ and $99.7\%$ error bars shows $\Delta_{\rm rms}$ and $R_{\rm med}$ values for the 11 brightest satellites selected using luminosities assigned with the \texttt{GRUMPY} model.

This figure shows that in the $\Delta_{\rm sph}-\Delta_{\rm rms}$ plane the MW satellite system is $\approx 2.5\sigma$ outlier in the case with no additional tidal disruption and $\approx 3\sigma$ outlier when such disruption is accounted for. As can be seen in Figure~\ref{fig:dsph_drms_hist}, the effect of the additional tidal disruption on $\Delta_{\rm sph}$ is rather minor {\it on average} and the change of the incidence is mainly due to the effect of tidal disruption on $\Delta_{\rm rms}$. However, the effect on individual systems can be significant and the sign of the effect can vary. 

Figure~\ref{fig:dsphk17} shows $\Delta_{\rm sph}(k)$ as a function of $k$ for the Caterpillar hosts 2 (right panel) and 17 (left panel). When we do not account for additional tidal disruption due to disk, the dispersion of satellite orbital poles is similar to the corresponding MW values for all $k\in [3,11]$ in Caterpillar host 17. This shows that dependence of $\Delta_{\rm sph}$ on $k$ measured for MW satellites is qualitatively similar to that expected in at least some of the $\Lambda$CDM haloes. However, when disk-induced disruption is accounted for the $\Delta_{\rm sph}(k)$ increases substantially and this statistic for the MW becomes $\approx 3\sigma$ away from the median values.  

On the other hand, the right panel of Figure~\ref{fig:dsphk17} shows that effect of accounting for disk disruption on $\Delta_{\rm sph}(k)$ has the opposite sign for the Caterpillar host 2. This accounting makes the distribution of orbital poles closer to that of the brightest MW satellites for this object.  

As we discussed in the next section, the magnitude of the disk tides on subhalo population is still theoretically uncertain. Results presented in this section show that this makes assessment of how common MW-like satellite systems are somewhat uncertain. Our results with and without accounting for disk-induced disruption likely bracket the range within which the true incidence lies.

\section{Discussion}
\label{sec:discussion}

Results presented in the previous section show that the MW satellite system is a $\approx 2-3\sigma$ outlier in terms of the thickness of its satellite plane and anisotropy of the orbital poles. The accuracy with which the incidence rate of MW-like satellite configurations can be estimated is limited by the uncertainty of the additional tidal disruption due to central baryonic disk. 

Our conclusions are rather different from the conclusions of \citet[][]{pawlowski_kroupa20}, who used Illustris TNG 100-1 simulations to quantify anisotropy of satellite systems around MW-sized galaxies. Specifically, these authors showed that a few percent of $\Lambda$CDM haloes have satellite planes with thickness comparable to that of the MW satellites. However, their results showed that MW is a far outlier in the $\Delta_{\rm sph}-\Delta_{\rm rms}$ plane with no systems in simulations matching properties of the MW satellite configuration. This led the authors to conclude that with the requirement that simulated satellite system has both $\Delta_{\rm rms}$ and $\Delta_{\rm sph}$ are as small as observed, the incidence of such systems in $\Lambda$CDM  ``essentially drops to zero.''

One difference between their approach and our study is that we focused on the anisotropy of the 11 brightest ``classical'' MW satellites, while most of the analysis of \citet{pawlowski_kroupa20} is focused on $k=7$ satellites for which the anisotropy is maximized. However, this difference is minor. We believe that the main difference is in the radial distribution of satellites used in their and our analyses. 

The TNG 100-1 hydro simulation used in \citet[][]{pawlowski_kroupa20} includes the tidal effects due to central baryon concentration in haloes. However, their results do not show the expected decrease in the mean concentration of satellite systems due to such tidal disruption. For example, their Figure 9 shows that distributions of $\Delta_{\rm rms}$ (denoted as $r_{\rm per}$ in that paper) in the hydro and dissipationless TNG 100 simulations are very similar with no significant increase in $\Delta_{\rm rms}$ that we see in our results (Figure~\ref{fig:rmed_drms_cdf}) when tidal effects of disk are modelled. 

The reason for this is likely the difference in resolution between TNG 100-1 and Caterpillar simulations. The dark matter particle masses in the hydro and dissipationless TNG 100-1 are  $m_{\rm DM}=7.47\times 10^6\, M_\odot$ and $m_{\rm DM}=8.86\times 10^6\, M_\odot$, respectively, while gravitational softening is $0.74$ kpc \citep[][]{Springel.etal.2018}. This means that haloes hosting typical ``classical'' MW satellites with $M_{\rm peak}\gtrsim 3-10\times 10^9\, M_\odot$ will be resolved with only $\sim 500-1000$ particles. This resolution is not sufficient to avoid numerical premature loss of subhaloes due to withering and artificial disruption \citep[][]{Green.etal.2021}. Indeed, we find that  in the ELVIS simulations, which have considerably better mass and force resolution ($m_{\rm DM}=1.9\times  10^5\, M_\odot$ and $\varepsilon=141$ pc), compared to the TNG 100-1 simulation the radial distribution of subhaloes is significantly affected by numerical effects for subhaloes resolved with $\lesssim 10^3$ particles (see Appendix~\ref{app:elvis_cat_rad_comp}). Furthermore, \citet{Webb.Bovy.2020} showed that artificial disruption significantly affects radial distribution of subhaloes even in modern state-of-the-art zoom-in hydrodynamical simulations of the MW haloes. 

In contrast, the Caterpillar halo simulations at the LX 14 resolution level used DM particle mass of $2.99\times 10^4\,M_\odot$ and graviational softening of $113$ pc in the zoom-in regions \citep[][]{Griffen.etal.2016}. The haloes of the classical satellites are therefore resolved with $\gtrsim 10^5$ particles and their evolution is followed with six times better force resolution. 

Qualitatively, our results are consistent with results of a recent study by \citet{sawala_etal22}. These authors have presented a similar analysis but using a model of orphan subhaloes to mitigate the effect of numerical disruption of subhaloes. They emphasized that inability of most previous simulations to properly model the radial distribution of subhaloes is the main 
reason why the incidence of the satellite systems as anisotropic as MW's was severely underestimated. 

A similar result was previously reported by \citet[][]{kang_etal05}, who analyzed the probability of getting planes as thick as observed in the MW and concluded that if satellites were distributed radially as dark matter, then the observed satellite configuration is expected to be reasonably common in $\Lambda$CDM. On the other hand, if satellites were distributed as subhaloes selected based on their current mass, the observed plane would be very rare. This finding is consistent with the results of \citet{sawala_etal22} and this study, which found that when satellite distribution is as radially concentrated as observed the thickness of the observed MW satellite plane is reproduced in $\approx 10\%$ of MW-sized $\Lambda$CDM haloes formed in dissipationless simulations.
However, when subhaloes are selected by their present mass, their radial distribution is much less concentrated than that of dark matter or observed satellites \citep[e.g.,][]{nagai_kravtsov05} and the observed thickness cannot be reproduced in most haloes due to the $\Delta_{\rm rms}-R_{\rm med}$ correlation discussed above in Section~\ref{sec:rmed_drms}.

Our results similarly show that in the Caterpillar simulations that reasonably reproduce the radial distribution of MW satellites \citep[see][]{Manwadkar.Kravtsov.2022}, the anisotropy of the MW spatial distribution and orbital poles is statistically consistent with the range of configurations in $\Lambda$CDM. The main difference from the analysis of \citet[][]{sawala_etal22} is that we do not use the orphan modelling, but rely on the inherent high resolution of the used Caterpillar suite to properly model tidal evolution of subhaloes in the relevant mass range (using tests reported in Section 4.4.4 and Figure 18 in \citealt{Manwadkar.Kravtsov.2022}). In addition, we explore effects of the tidal disruption due to disk not considered by \citet{sawala_etal22}. In particular, we show that these effects make distribution less radially concentrated and less anisotropic. Ignoring these effects thus overestimates the incidence of MW-like anisotropic satellite systems.  

Using a model that takes into account effects of disk on subhalo survival, we estimate that the Milky Way is a $\approx 2.5-3\sigma$ outlier in the expected $\Lambda$CDM distribution estimated using the Caterpillar suite. This significance is somewhat uncertain due to uncertainties in the disk disruption model. 

Specifically, as was shown recently by \citet[][]{Webb.Bovy.2020} and \citet{Green.etal.2022}, modern zoom-in hydro simulations of the MW-sized haloes significantly overestimate disruption of subhaloes due to the presence of central disk. The effect of such disruption in our study thus may also be overestimated because disruption model of \citet{Nadler.etal.2018} that we use was calibrated on such zoom-in simulations from the FIRE suite.
This implies that the incidence of the MW-like satellite planes is likely between our estimates with and without accounting for disk disruption. 
Overall, therefore, {\it MW satellite system is somewhat unusual but is statistically consistent with $\Lambda$CDM.}

In agreement with \citet{Ahmed.etal.2017}, our results show that the common assumption that plane of the satellites problem of $\Lambda$CDM and is not sensitive to baryonic effects is not quite correct as tidal disruption associated with the presence of the central baryonic disk does affect the estimate of the incidence rate for MW-like satellite systems and thereby conclusions about whether anisotropy of the MW satellite distribution is consistent with $\Lambda$CDM. 

It is worth noting that the Milky Way is somewhat unusual not only in the anisotropy of its satellite system, but also in its environment. For example, the Local Group is embedded in the flattened galaxy distribution called the Local Sheet, which itself is a part of a larger structure called the Local Supercluster. The Local Sheet is overabundant in massive galaxies, which makes it a $\approx 2-3\sigma$ outlier among regions of similar overdensity in $\Lambda$CDM \citep[][]{Neuzil.etal.2020}. 

Furthermore, the correlation of the satellite plane orientations with the orientation of the tidal field in the Local Supercluster region \citep[][]{Libeskind.etal.2015} indicates that the formation of such anisotropic satellite system is likely shaped by our specific cosmic neighborhood. 
The comprehensive assessment of probability of the anisotropy of the satellite distribution in the MW and other bright nearby galaxies will thus require both careful matching of 
statistical measures used to quantify observed anisotropy and proper exploration and modelling of the effects of our local cosmic environment on the formation of satellite systems. Interestingly, \citet{Dupuy.etal.2022} do find that in constrained simulations of the Local Group, in which  our observed local environment is matched, satellites tend to accrete onto MW and M31 along direction of the slowest collapse of the Local Sheet. However, they also find that non-linear dynamics following accretion scrambles this correlation.

\section{Summary and conclusions}
\label{sec:summary}

In this study, using a suite of zoom-in, high-resolution, dissipationless simulations of the MW-sized haloes, we examined the incidence of satellite systems as anisotropic as that of the MW in terms of their spatial and orbital distribution. We re-confirmed that this incidence is sensitive to the radial distribution of satellites \citep[][and see Figure~\ref{fig:drms_rmed} in this paper]{kang_etal05,zentner_etal05}, which can be affected by artificial subhalo disruption due to numerical effects and due to disruption induced by the tidal force of the central baryonic disk. These effects on the subhalo population are uncertain because they depend on the proper modelling of the MW disk evolution \citep[see, e.g., Appendix A in][]{Bose.etal.2020} and require very high resolution to minimize effects of artificial disruption \citep[see][]{Webb.Bovy.2020,Green.etal.2022}. 

In this study, we try to control for effects of artificial disruption by using the large Caterpillar suite of simulations with the resolution sufficiently high to reliably model radial distribution of the brightest satellites (see Section~\ref{sec:modelling} and Appendix~\ref{app:elvis_cat_rad_comp}). We bracket effects of the tidal disruption due to disk by estimating the incidence of the MW-like systems without accounting for such effects and with accounting for them with the model of \citet[][]{Nadler.etal.2018}, which likely overestimates the effect \citep[see Section 5 of][]{Webb.Bovy.2020}.

Our main results and conclusions can be summarized as follows. 

\begin{itemize}
\item[(i)] When we do not account for the disk-induced disruption of subhaloes, we find that $\approx 10\%$ of the satellite systems around $\Lambda$CDM MW-sized haloes are more spatially flattened than the MW bright satellites (see Figure~\ref{fig:drms_rmed_hist}). When both flattening and orbital pole anisotropy are considered, the MW system is a $\approx 2-3\sigma$ outlier among the $\Lambda$CDM MW-sized haloes (see Figure~\ref{fig:dsph_drms_hist}).\\
\item[(ii)] When the disk induced tidal disruption is accounted for using the model of \citet{Nadler.etal.2018} calibrated with FIRE-2 simulations, only $\approx 1\%$ of the $\Lambda$CDM haloes host satellite systems as flattened as in the MW. MW is $\approx 3\sigma$ outlier in the case when both flattening and orbital pole anisotropy are considered (see Figure~\ref{fig:dsph_drms_hist}). \\
\item[(iii)] We show that at least one halo out of 32 MW-sized haloes in the Caterpillar suite reproduces the maximum orbital pole dispersion statistic, $\Delta_{\rm sph}(k)$, as a function of the $k$ number of satellites (see Figure~\ref{fig:dsphk17}).\\ 
\item[(iv)] Sensitivity of the incidence of the MW-like satellite systems to baryonic modelling of the central disk pointed out by \citet{Ahmed.etal.2017} and confirmed by our results implies that careful modelling of these effects, while controlling for resolution effects on subhalo disruption, will be required for reliable estimate of such incidence in the $\Lambda$CDM model.\\
\item[(v)] Assuming that our models with and without taking into account additional disruption of subhaloes by the central disk bracket the effects of such disruption, we estimate that the MW satellite system is a $\approx 2-3\sigma$ outlier among population of the MW-sized hosts in terms of the flatness and orbital anisotropy of its population of bright satellites. 

\end{itemize}

Overall, our results indicate that the {\it MW satellite system is somewhat unusual but is statistically consistent with $\Lambda$CDM model}, in general agreement with results and conclusions of recent studies by \citet{Samuel.etal.2021}, \citet[][]{sawala_etal22}, and \citet{Foerster.etal.2022}.

\section*{Acknowledgements}
We are grateful to Phil Mansfield for many useful discussions. 
We would like to thank Shea Garrison-Kimmel and Michael Boylan-Kolchin for providing halo tracks of the ELVIS simulations.
We thank Alexander Ji and the Caterpillar collaboration for providing halo tracks of the Caterpillar simulations used in this study.
This work was supported by the National Science Foundation grants AST-1714658 and AST-1911111 and NASA ATP grant 80NSSC20K0512.
Analyses presented in this paper were greatly aided by the following free software packages: {\tt NumPy} \citep{numpy_ndarray}, {\tt SciPy} \citep{scipy}, {\tt Matplotlib} \citep{matplotlib}, and \href{https://github.com/}{\tt GitHub}. We have also used the Astrophysics Data Service (\href{http://adsabs.harvard.edu/abstract_service.html}{\tt ADS}) and \href{https://arxiv.org}{\tt arXiv} preprint repository extensively during this project and the writing of the paper.

\section*{Data Availability}

Halo catalogs from the ELVIS simulations can be obtained upon request from Shea Garrison-Kimmel and Michael Boylan-Kolchin, while the data for the Caterpillar simulations is available at \href{https://www.caterpillarproject.org/}{\tt https://www.caterpillarproject.org/}. The \texttt{GRUMPY} model pipeline is available at \url{https://github.com/kibokov/GRUMPY}. The data used in the plots within this article are available on reasonable request to the corresponding author.

\bibliographystyle{mnras}
\bibliography{refs}



\appendix

\section{Cumulative radial distributions of subhaloes in the ELVIS and Caterpillar simulations}
\label{app:elvis_cat_rad_comp}

\begin{figure*}
    \centering
    \includegraphics[width=0.99\textwidth]{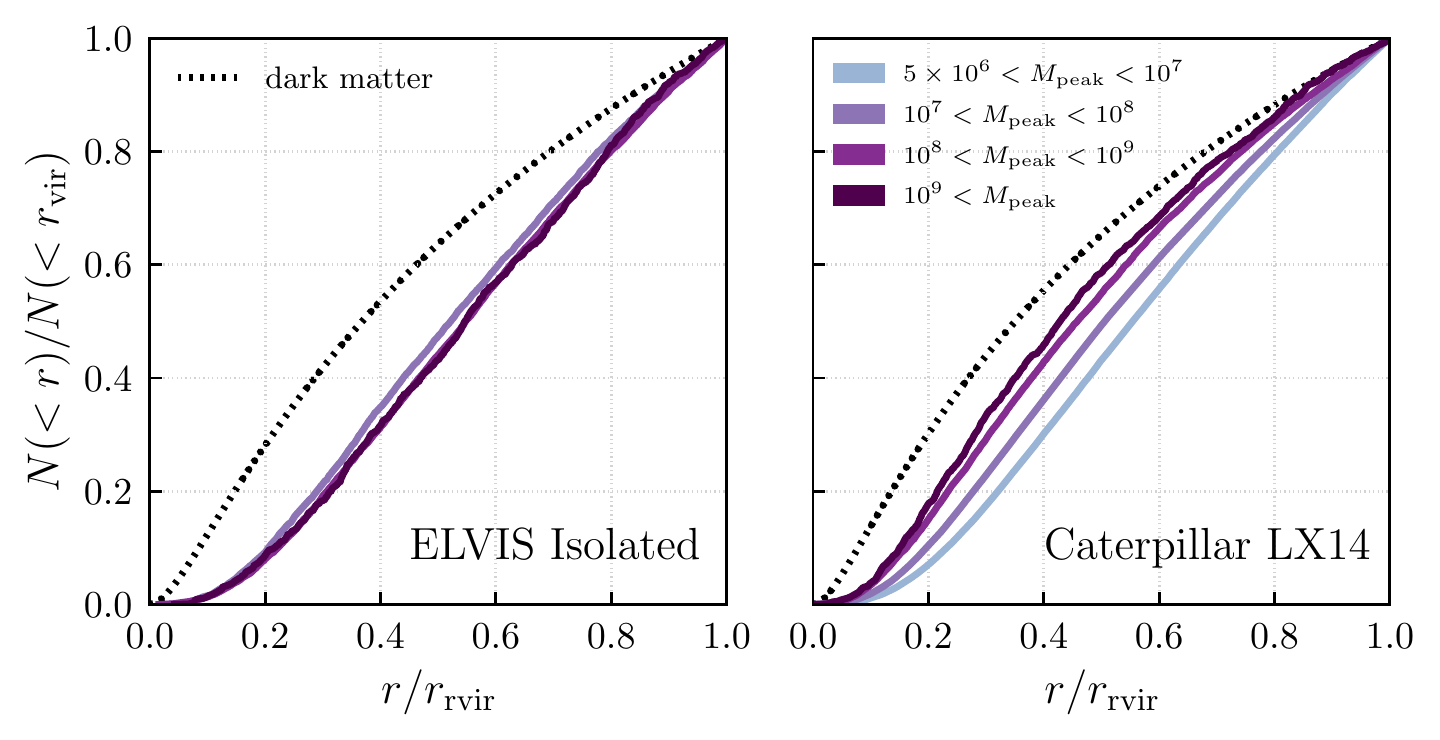}
    \caption{The cumulative radial distribution of the number of subhaloes around  MW-sized hosts (solid lines) compared to the dark matter cumulative mass distribution (dotted lines) in the ELVIS (left panel) and Caterpillar (right panel) suites. Solid lines of different colours show radial distributions of subhaloes in different $M_{\rm peak}$ ranges, as indicated in the legend. Here $M_{\rm peak}$ is the maximum mass reached by a subhalo throughout its evolution. }
    \label{fig:elvis_cat_pro_comp}
\end{figure*}

\begin{figure}
    \centering
    \includegraphics[width=0.49\textwidth]{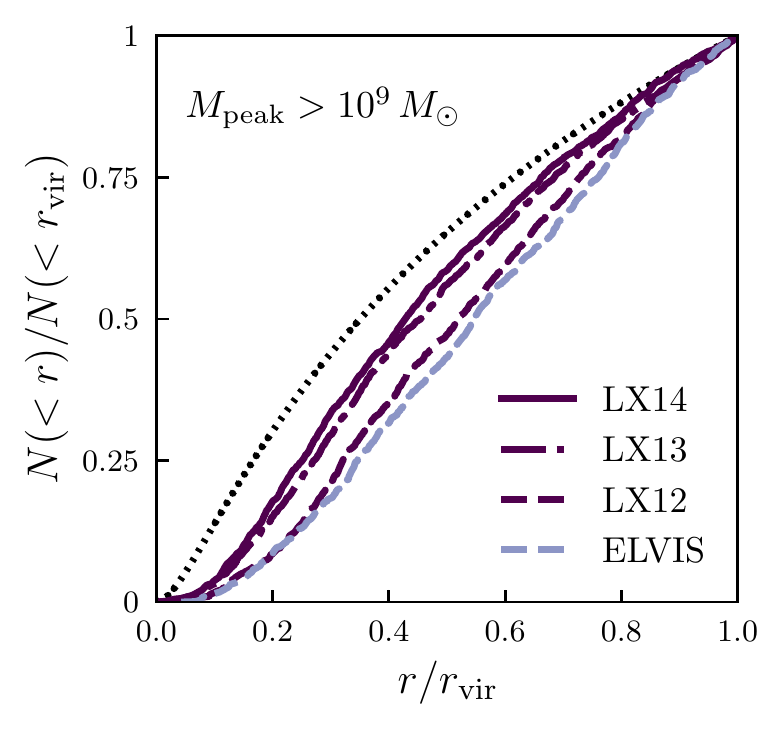}
    \caption{Radial disributions of the most massive subhaloes of $M_{\rm peak}>10^9\,\Msun$ in the ELVIS MW-sized haloes (light blue dashed line) and in the Caterpillar haloes simulated at three different resolutions (see Section~\ref{sec:caterpillar} for details), shown by dark blue dashed, dot-dashed and solid lines for the resolution levels LX12, LX13, and LX14, correspondingly. The dotted line shows the average dark matter radial mass distribution in the Caterpillar hosts.}
    \label{fig:elvis_cat_pro_comp_res}
\end{figure}

We compare cumulative radial distributions of subhaloes in the ELVIS and Caterpillar simulation suites (described in Section~\ref{sec:modelling}) in Figures~\ref{fig:elvis_cat_pro_comp} and \ref{fig:elvis_cat_pro_comp_res}. 

Figure~\ref{fig:elvis_cat_pro_comp} shows the cumulative radial distribution of the number of subhaloes around  MW-sized hosts compared to the average dark matter cumulative mass distribution (shown by the dotted lines) in the ELVIS (left panel)  and Caterpillar (right panel) simulation suites for subhaloes of different $M_{\rm peak}$ ranges. The figure shows that radial distribution of subhaloes in the ELVIS simulations does not depend on $M_{\rm peak}$. The lowest mass subhaloes used in the figure should definitely be significantly affected by the artificial disruption \citep[see][]{vandenBosch.etal.2018,vandenBosch.Ogiya.2018,Webb.Bovy.2020,Green.etal.2021} that should make the radial distribution less concentrated. Larger mass subhaloes are better resolved and thus should be less and less subject to numerical disruption effects as subhalo mass increases \citep[see, e.g., tests and discussion in Section 5 of][]{Webb.Bovy.2020}. It is thus quite surprising not to see any trend of the radial distribution with subhalo mass in the ELVIS suite.  

Radial distributions of subhaloes in the Caterpillar suite, on the other hand, do become more centrally concentrated with increasing subhalo mass, as expected for subhaloes less affected by numerical disruption. Moreover, Figure~\ref{fig:elvis_cat_pro_comp_res} compares radial distributions of the most massive subhaloes of $M_{\rm peak}>10^9\,\Msun$ in the ELVIS MW-sized haloes and in the Caterpillar haloes simulated at three different resolutions (see Section~\ref{sec:caterpillar} for details). We can see that the radial distribution of the ELVIS subhaloes is similar to that of the Caterpillar subhaloes at the lowest (LX12) resolution limit. The radial distribution of the Caterpillar subhaloes becomes more concentrated with increasing resolution, although distributions at the LX13 and LX14 resolution levels are similar, indicating convergence for subhaloes of this mass.  

Although a more detailed study of the subhalo disruption in these simulations would be warranted, overall results shown in these figures indicate that subhaloes in the ELVIS simulations are largely in the regime where artificial disruption affects subhaloes of all relevant masses, while in the Caterpillar suite the most massive subhaloes used in this study are little affected by the artificial disruption and a negligible fraction of these subhaloes are lost to ``withering'' or crossing the lowest resolution threshold \citep[see][]{Green.etal.2021}. These conclusions are consistent with the trends seen in the numerical tests reported in Section 5 of \citet[][]{Webb.Bovy.2020}.

\bsp	
\label{lastpage}
\end{document}